\documentclass[12pt,letterpaper,usenames,dvipsnames]{article}
\usepackage{jheppub}

\usepackage{tabstackengine}
\setstacktabbedgap{1ex}

\allowdisplaybreaks[2]  
\usepackage{longtable}
\usepackage{pifont}
\usepackage{bm} 
\usepackage{mathrsfs}   
\usepackage{ytableau}   
       
\usepackage{float}
\usepackage{hhline}
\usepackage{mathtools}
\usepackage[graphicx]{realboxes}
\usepackage{xspace}
\definecolor{aqua}{rgb}{0, 1.0, 1.0}
\definecolor{fuschia}{rgb}{1.0, 0, 1.0}
\definecolor{gray}{rgb}{0.502, 0.502, 0.502}
\definecolor{lime}{rgb}{0, 1.0, 0}
\definecolor{maroon}{rgb}{0.502, 0, 0}
\definecolor{navy}{rgb}{0, 0, 0.502}
\definecolor{olive}{rgb}{0.502, 0.502, 0}
\definecolor{purple}{rgb}{0.502, 0, 0.502}
\definecolor{silver}{rgb}{0.753, 0.753, 0.753}
\definecolor{teal}{rgb}{0, 0.502, 0.502}

\newcommand{\sdim}{{\mathrm {sdim}\xspace}}

\graphicspath{{figures/}}	 
\usepackage{tikz}
\usetikzlibrary{arrows,calc,snakes,shapes,shapes.arrows,
decorations.markings,decorations.pathmorphing}
\newcounter{sarrow}

     \tikzstyle{gr}=[draw,circle,green!50!black,fill=green!50!black,scale=.6]
     \tikzstyle{Bl}=[draw,circle,blue,scale=.7]
     \tikzstyle{R}=[draw,circle,fill=red,scale=.7]
     \tikzstyle{bl}=[draw,circle,fill=black,scale=.2]
     \tikzstyle{bbc}=[draw,circle,fill=black,scale=.75]
     \tikzset{->-/.style={decoration={markings, mark=at position #1 with {\arrow{>}}},postaction={decorate}}}
 
\usepackage{array} 
\usepackage{multirow} 
\usepackage{colortbl} 
\usepackage{arydshln} 
\usepackage{stfloats}  
\usepackage{cellspace}
\usepackage{xcolor}   

\usepackage{url} 
\usepackage{verbatim} 

\def\bar{\overline}

\newcommand{\beq}{\begin{equation}}
\newcommand{\eeq}{\end{equation}}
\newcommand{\bpm}{\begin{pmatrix}}
\newcommand{\epm}{\end{pmatrix}}
\newcommand{\bpmat}{\begin{pmatrix}}
\newcommand{\epmat}{\end{pmatrix}}
\newcommand{\bsmat}{\begin{smallmatrix}}
\newcommand{\esmat}{\end{smallmatrix}}

\def\^{\wedge}

\def\Sp{{\rm\, Sp}}

\def\Z{\mathbb{Z}}

\def\cH{{\mathcal H}}

\def\cN{{\mathcal{N}}}

\def\-{{\text{-}}}

\preprint{{\flushright
UTTG--01--19\\
}}

\title{On the BPS Spectrum of the Rank-1 Minahan-Nemeschansky Theories}
\author[1]{Jacques Distler,}
\author[1]{Mario Martone,}
\author[2]{Andrew Neitzke}
\affiliation[1]{University of Texas, Austin, Physics Department, Austin TX 78712}
\emailAdd{distler@golem.ph.utexas.edu}
\emailAdd{mariomartone@utexas.edu}
\affiliation[2]{University of Texas, Austin, Mathematics Department, Austin TX 78712}
\emailAdd{neitzke@math.utexas.edu}

\abstract{The F-theory realization of the rank-1 Minahan Nemeschansky (MN) $E_6$, $E_7$ and $E_8$ theories leads to a description of the BPS states on the 
Coulomb branch in terms of Type IIB $(p,q)$-string networks. 
Subject to a simple ansatz for the types of networks which can occur,
we study the representations of
the flavor symmetry group which occur in the BPS spectrum. 
The results we find for the $E_6$ and $E_7$ theories are in perfect agreement with previous calculations by other methods (in particular, we find that arbitrarily large representations occur),
but our scheme is easier to implement and more computationally efficient.
The string network picture also gives a possible explanation of the experimental observation that in rank-1 MN theories, BPS states whose charge is $n$ times a primitive charge occur with BPS index divisible by $(-1)^{n+1} n$.}

\begin{document}
\maketitle


\section{Introduction}

{\bf BPS one-particle states in $\cN=2$ theories.}
The spectrum of $\frac12$-BPS one-particle states on the Coulomb branch 
belongs to the short list
of exactly computable observables of an $\cN=2$ supersymmetric field theory.
As such it has been a basic object of study at least since the work
of Seiberg and Witten \cite{Seiberg:1994aj,Seiberg:1994rs}, where the spectrum for the pure $SU(2)$ theory was computed. 
Analogous problems in $\cN=(2,2)$ supersymmetric theories
in two dimensions had been considered earlier, e.g. in \cite{Cecotti:1993rm} 
which proposes to use the BPS spectrum as the basis for a classification
of all such theories admitting a massive deformation --- loosely speaking,
to show that the theory can be reconstructed from its BPS spectrum.
It remains to be seen whether such a program has a chance to work in
four dimensions, but at least the BPS spectrum contains a lot of 
information about the theory.

By now there is a large body of work on BPS
spectra in various four-dimensional $\cN=2$ field theories, from many points of view. We cannot give a complete
review here, but just mention some of the lines of development and a few works in each line:
\begin{itemize}
\item quiver quantum mechanics and related ideas, e.g. \cite{Denef2002,Alim2011,Manschot2014,Hori2014,Kim2015,Cordova2015},
\item wall-crossing methods and consistency constraints, e.g. \cite{Ferrari:1996sv,Denef2007,ks1,Gaiotto2008,Chuang2008,Cecotti2009,Gaiotto2012,Galakhov:2013oja,Longhi2016,Gabella2017} (see \cite{Pioline2011} for a review of this approach as of 2011, containing many more references),
\item geometric realizations such as string networks, or more generally split attractor flows on the Coulomb branch, e.g. \cite{Klemm1996,Schwarz:1996bh,Mikhailov:1998bx,DeWolfe:1998zf,DeWolfe:1998eu,Aharony:1996xr,Denef2001}.
\end{itemize}

These lines of development have considerable overlap, and each
has informed the others, leading to a rich,
tightly constrained and consistent picture.
Nevertheless, the list of theories where the spectrum has been
completely and explicitly described is still short, and surprises continue
to be discovered.

\medskip

{\bf String networks in F-theory.}
In this paper we explore what can be learned by applying 
the F-theory perspective to a
 rank-$1$ Minahan-Nemeschansky theory of type $E_{6}$, $E_7$ or $E_8$ \cite{Minahan:1996fg,Minahan:1996cj}. This theory is the $\mathcal{N}=2$ SCFT realized on the world-volume of a single D3-brane probing an exceptional 7-brane in F-theory \cite{Sen:1996vd,Dasgupta:1996ij,Johansen:1996am,Gaberdiel:1997ud,Aharony:2007dj}  and it has a $1$-dimensional Coulomb branch, corresponding to moving the D3-brane in the plane transverse to the 7-brane. The $\frac12$-BPS particles of the 4D theory can be realized as $(p,q)$-strings which stretch between the D3-brane and the exceptional 7-brane \cite{Aharony:1996xr,Schwarz:1996bh}.

These $(p,q)$-strings are challenging to study directly, because the physics of strings ending on the exceptional 7-brane is complicated. To get around this problem, we imagine mass-deforming the exceptional 7-brane into component D7-branes (in F-theory terms, deforming an $IV^*$, $III^*$ or $II^*$ singularity into a collection of $I_1$ singularities). If we are far out on the Coulomb branch (i.e., if the distance to the D3-brane is much greater than the separation of the D7-branes), this deformation does not affect the spectrum of BPS states: 
the spectrum is still that of the conformal theory.\footnote{We emphasize that 
``the spectrum of the conformal theory'' is well defined: for conformal theories
with a $1$-dimensional Coulomb branch,
there are no wall-crossing phenomena to worry about.} However,
after the deformation, the way the BPS states are realized is different; we still have a single string stretching out to the D3-brane, but at the 7-brane end, the string now splits into a network of string junctions, with prongs ultimately ending on the component D7-branes.

\medskip

{\bf Our ansatz.}
To study these string junctions, we follow a strategy pioneered in
\cite{DeWolfe:1998zf,DeWolfe:1998eu}. 
After 
lifting to $F$-theory, BPS string junctions can be identified
with holomorphic curves in a torus fibration $X$ over the Coulomb branch.
Any holomorphic curve must have nonnegative genus, and any two
holomorphic curves must have nonnegative intersection number.
On the other hand, one can read off the flavor and electromagnetic
charges of the BPS states from the relative homology classes 
of the corresponding holomorphic curves (see \cite{Grassi2014} for a 
detailed account of this point.)
Combining these facts gives
constraints on the possible charges of BPS states. 
Next we make a simple ansatz: we assume that \emph{every charge which is allowed by 
these constraints is actually realized as the charge of a BPS state}.

This kind of ansatz has been used before, e.g. in \cite{Mikhailov:1998bx,DeWolfe:1998eu} for $SU(2)$ theories with fundamental flavors, and \cite{Halverson2016} for certain Argyres-Douglas theories.
In all these cases this ansatz turned out to give the correct BPS spectrum.
Moreover the spectrum turned out to be fairly simple: for example,
in the $SU(2)$ $N_f=4$ theory, 
the only representations of the $Spin(8)$ flavor symmetry
which occur are the ${\bf 1}, {\bf 8}_s, {\bf 8}_c, {\bf 8}_v$.
In contrast, in the Minahan-Nemeschansky theories which we consider,
our ansatz leads to a much more complicated result:
for states with
electromagnetic charge $n(p,q)$, larger and larger flavor representations occur in the spectrum as we increase $n$.
To take a concrete example, in the $E_6$ MN theory, we find that
for states with electromagnetic charge $(6,0)$ the possible
$E_6$ representations are ${\bf \overline{5824}}$, ${\bf 2430}$,
${\bf 2925}$, ${\bf 650}$, ${\bf 78}$, ${\bf 1}$; 
for other examples see Tables \ref{E6n10}, 
\ref{E6npq}, \ref{E7n10}, \ref{E7npq}, \ref{E8n10}, \ref{E8different},
\ref{E8npq} below.

The prediction that arbitrarily large representations should appear 
in the BPS spectrum runs counter to one's experience with simpler
theories, and might lead one to be skeptical of our ansatz.
However, we find reason for optimism:
the BPS spectrum for some electromagnetic charges 
in the $E_6$ and $E_7$ MN theories have recently been calculated
by a different method \cite{Gaiotto2012,Hollands:2016kgm,E7bps},
and the list of representations we obtain using our ansatz 
precisely matches what was computed there! This agreement leads 
us to conjecture that our ansatz is actually correct.

Supposing that it is indeed correct,
this string-network method gives a quite efficient way of computing the charges
which can occur in the BPS spectrum. In particular, the string-network method
is much faster than the spectral-network methods applied to MN theories in \cite{Hollands:2016kgm,E7bps},
which (so far) are practical for a charge $n(p,q)$ only when all of $n$, $p$, $q$
are small.

\medskip

{\bf BPS multiplicities.}
The string-network method (at least in our hands) 
does not directly give a recipe for the BPS \emph{multiplicities}: it tells
us only which representations occur, not how many times they occur. 
One might hope that the indices can be somehow extracted
from a closer look at the string networks or their associated holomorphic curves.
We make some preliminary exploration along these lines.

In \cite{Hollands:2016kgm,E7bps}
it was observed that the indices computed there have
an unexpected divisibility property: for BPS particles of electromagnetic 
charge $n(p,q)$, where $\gcd(p,q) = 1$, 
the BPS index is always a positive integer multiple of $(-1)^{n+1} n$.
The string-network picture of the BPS particles 
offers a possible explanation of this phenomenon, as
follows. A state of charge $n(p,q)$ comes from an M2-brane wrapping 
a holomorphic curve whose boundary
is homologous to an $n(p,q)$-cycle on the torus fiber at the D3-brane.
If the boundary is a union of disjoint, simple, essential closed curves, then to realize
the class of an $n(p,q)$-cycle requires exactly $n$ boundary components (as
pointed out in this context in \cite{DeWolfe:1998eu}).
Plausibly (see \S\ref{purity}), the quantization of M2-branes wrapping holomorphic curves with $n$ boundary components 
produces a reduced Hilbert space of the form $V_n \otimes W$ 
where $V_n$ is a ``universal'' multiplet of spin $\frac{n}{2}$.\footnote{A stronger hypothesis would be that $W$ is a sum of copies of the spin-$0$ representation, so that the full reduced Hilbert space would be a sum of spin-$\frac{n}{2}$ representations. This hypothesis was proposed in \cite{Hollands:2016kgm} where
it was called ``spin purity.'' While spin purity could be true, 
all the evidence so far
is also consistent with the weaker factorization hypothesis above.}
This has been observed
before for $n=1,2$ which give rise, respectively, to BPS hypermultiplets and vector multiplets \cite{Klemm1996,Henningson:1997hy,Mikhailov:1997jv}. If it 
is true for arbitrary $n$, then
this would explain the divisibility phenomenon in rank $1$ Minahan-Nemeschansky
theories.

Now what about the actual indices?
We can report
one encouraging experimental result, obtained by comparing
the results of \cite{Hollands:2016kgm,E7bps} to the string-network
picture: it appears that the 
BPS index is precisely $(-1)^{n+1} n$ --- the contribution from a \emph{single} spin-$\frac{n}{2}$ multiplet --- if and only if the corresponding holomorphic
curve has genus $0$. However, when the genus is greater than $0$, the story becomes
more complicated, and despite some intriguing regularities\footnote{For instance, for $g=1$ we find only indices $2$ and $3$ occur.}, we do not have a general rule for determining the index.

\medskip

{\bf Future directions.} Here are a few questions which this work suggests:
\begin{itemize}
	\item \emph{Why} is our ansatz correct? The construction of string junctions
	is in principle highly constrained: each string has to follow a gradient
	trajectory on the $u$-plane for an appropriate central charge function.
	Nevertheless it seems that anything that is allowed by our simple 
	topological constraints actually does occur.
	It would be very nice to give a direct construction of the required
	junctions.

	\item Can we carry out the quantization of the M2-branes directly,
	to establish that indeed M2-branes wrapping holomorphic curves
with $n$ boundary components only produce states in spin-$\frac{n}{2}$ multiplets?

	\item Can we give rules for determining the precise BPS indices
	arising from M2-branes with genus $g \neq 0$?

	\item Can we use F-theory technology to study BPS states in the higher-rank
	MN theories, obtained by considering $r>1$ D3-branes probing a single
	exceptional 7-brane? (One remark we can make immediately is that in 
	this case the relation between the number of 
	boundary components and the charge will be a bit different: see Section \ref{geometry_and_properties_of_the_bps_states} below.)
\end{itemize}

\section{Review of existing literature}

There is a tremendously rich literature discussing BPS states in $\cN=2$ theories and how they arise in various string realizations. We will briefly review the key parts which are needed in our analysis of BPS states in this section; for more details we refer the reader to the original literature.

\subsection{String junctions and BPS states}\label{string_junctions_and_bps_states}

String junctions \cite{Aharony:1996xr,Schwarz:1996bh} give a useful way of realizing the spectrum of BPS states of 4-dimensional $\mathcal{N}=2$ theories which can be realized on the world-volume of a Type IIB 
D3-brane probing orientifold singularities \cite{Fayyazuddin:1997cz,Bergman:1998br,Bergman:1998ej}.

In what follows it will be convenient to reinterpret the string junctions
in terms of holomorphic curves. Here we just briefly 
recall how that picture works; see e.g. \cite{Grassi2014} for a more 
detailed account of the relevant geometry.
The Coulomb branch of the $\mathcal{N}=2$ theory is a $1$-dimensional complex space,
parametrized by the position of the D3-brane in the transverse space 
to the orientifold singularity. The F-theory torus fibers over the Coulomb
branch, giving an elliptically fibered complex surface $X$, which is
in fact hyperk\"ahler (one may think of it as like a local patch of a K3 surface).
BPS states in the theory at $u$ correspond to open membranes $D \subset X$ 
whose boundary lies in the fiber $X_u$ over $u$; the BPS condition requires that
$D$ is actually \textit{holomorphic} in one of the complex structures
on $X$.

The electromagnetic and flavor charges 
of the BPS state are determined by the relative homology class
$[D] \in H_2(X;X_u)$. In particular, the boundary $\partial D$ is a $1$-cycle
on $X_u$; we write its homology class $[\partial D] \in H_1(X_u)$
as $n(p,q)$, where we take $p$ and $q$ relatively
prime but not necessarily positive; following the string junction literature 
we sometimes refer to this as the \textit{asymptotic charge}
of the corresponding string network; it is also the electromagnetic charge
of the corresponding BPS state.

To understand why BPS states are related to holomorphic curves, we recall how the IIB picture can be lifted to M-theory. We start with the F-theory picture, then 
compactify on a circle and T-dualize.
Then the elliptic fiber of $X$ becomes part of the eleven-dimensional M-theory background. In this lift the D3-brane probe lifts to a M5-brane on the torus
fiber, and $n(p,q)$-strings stretched between the D3 probe and the D7 lift to M2-branes ending on the M5 and wrapping the $n(p,q)$-cycle of the torus fibers. 
Thus string junctions in type IIB lift to membranes wrapping complex curves in the total space of $X$.

At any rate, in this language, 
the question of whether BPS states exist with given charges
gets translated to the question whether a given class in $H_2(X;X_u)$ contains a holomorphic representative. This is in general a difficult problem, but we can at
least give some necessary conditions.
First, the self-intersection of a holomorphic curve $\mathcal{J}$ with genus $g$ and $b$ boundary components in $X$ is (using the fact that $X$ has trivial canonical bundle):
\begin{equation}
\#(\mathcal{J}\cdot\mathcal{J}) = -\chi(\mathcal{J}) = 2g-2+b,
\label{holcur}\end{equation}
where $\chi(\mathcal{J})$ is the Euler characteristic of $\mathcal{J}$. 
Writing $J$ for the homology class $[\mathcal J]$,
the holomorphy of $\mathcal{J}$ now descends to some constraints on $J$, as follows.
Following \cite{DeWolfe:1998eu}, $b$ can be identified with $n$, the greatest common divisor of the electromagnetic charges of $J$. Since $g\geq 0$, \eqref{holcur} then implies
\begin{equation}
(J, J)\geq-2+n,
\label{selfint}\end{equation}
where now $(\;,\;)$ denotes the intersection pairing.\footnote{We are taking
the intersection here between \textit{relative} 
homology cycles; fortunately this number
is indeed well defined in our context for cycles whose boundaries are homologous.}
While \eqref{selfint} is clearly necessary, not every class $J$ obeying
\eqref{selfint} admits a holomorphic representative. Another necessary condition
is that if $J$ and $J'$ are distinct and have holomorphic representatives then
they have nonnegative mutual intersection,\footnote{There is a tricky point here: this constraint
applies to curves which are holomorphic in the same complex structure on $X$, but
BPS states with different central charges are generally holomorphic in 
different complex structures on $X$.
Indeed the complex structure is determined by the phase of the central charge. In our case, when
the D3-branes are very far from the D7-branes, the central charge is dominated by the contribution
from the asymptotic charges; thus two BPS states which have the same asymptotic charge are holomorphic
in complex structures which can be made arbitrarily close, which is enough to ensure
that the intersection constraint holds.}
\begin{equation}
(J,J')\geq 0.
\label{mutualint}\end{equation}

Below we will work out concretely the set of charges $J$ compatible with 
the conditions\footnote{You might worry that there is an ambiguity here. Once we have \emph{at least one} $J$ which we know to be BPS, we can apply \eqref{mutualint} to any new candidate BPS junction, $J'$, to determine whether it is also BPS. But we need to start somewhere. We will assume that, whenever we find a charge $J$ which saturates \eqref{selfint} for $n=1$,  then $J$ has a holomorphic representative. This is sufficient to bootstrap the rest of the BPS spectrum.} \eqref{selfint} and \eqref{mutualint}.
Comparing our results with the BPS spectra computed in \cite{Hollands:2016kgm,E7bps}, we will find that \eqref{selfint} and \eqref{mutualint} appear to be not only necessary but also sufficient
for the existence of a BPS state! Encouraged by this striking agreement, we will then go on to use \eqref{selfint} and \eqref{mutualint} to study the BPS spectrum for the $E_8$ MN theory. To our knowledge, this is the first time that an analysis of the BPS spectrum of this theory has been performed, though it is not the first time that it is noticed that \eqref{selfint} and \eqref{mutualint} compute the complete BPS spectrum of an $\cN=2$ theory. In \cite{Mikhailov:1998bx,DeWolfe:1998eu} it was in fact already pointed out that that these two constraints are sufficient to give the well-known, albeit far simpler, BPS spectrum of the $\cN =2$ SU(2) theory with $0\leq N_f\leq 4$ flavors, and in \cite{Halverson2016}
the same was done for some Argyres-Douglas theories.

\subsection{Intersection pairing and flavor symmetry}\label{intersection_form_on_the_string_junctions_and_flavor_symmetry}

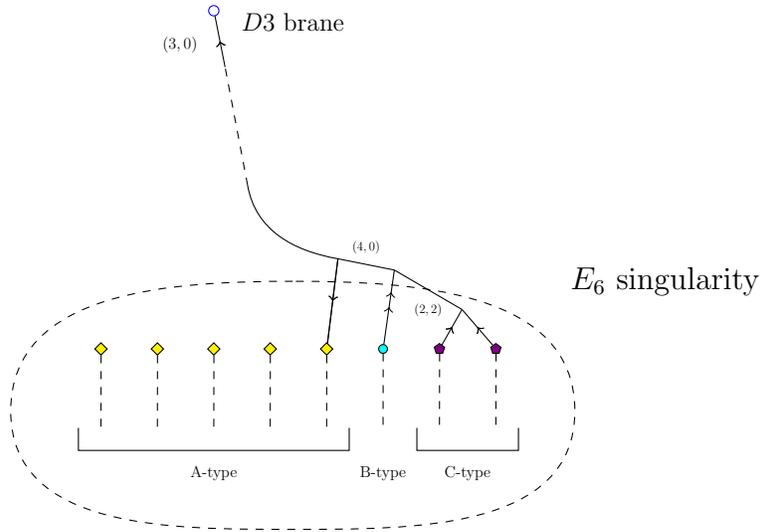
\begin{figure}[tbp]
\centering
\begin{tikzpicture}[decoration={markings,
mark=at position .5 with {\arrow{>}}}]
\begin{scope}[scale=1.5]
\node[Bl,scale=.5] (or1) at (8,3) {};
\node[scale=.8] at (8.7,2.9) {$D3$ brane};
\node[draw,diamond,fill=yellow,scale=.3] (A1) at (7,0) {}; 
\draw[dashed] (A1) -- (7,-.7);
\node[draw,diamond,fill=yellow,scale=.3] (A2) at (7.5,0) {}; 
\draw[dashed] (A2) -- (7.5,-.7);
\node[draw,diamond,fill=yellow,scale=.3] (A3) at (8,0) {}; 
\draw[dashed] (A3) -- (8,-.7);
\node[draw,diamond,fill=yellow,scale=.3] (A4) at (8.5,0) {}; 
\draw[dashed] (A4) -- (8.5,-.7);
\node[draw,diamond,fill=yellow,scale=.3] (A5) at (9,0) {};
\draw[dashed] (A5) -- (9,-.7);
\node[draw,circle,fill=aqua,scale=.3] (B) at (9.5,0) {};
\draw[dashed] (B) -- (9.5,-.7);
\node[draw,regular polygon,regular polygon sides=5,fill=purple,scale=.3] (C1) at (10,0) {}; 
\draw[dashed] (C1) -- (10,-.7);
\node[draw,regular polygon,regular polygon sides=5,fill=purple,scale=.3] (C2) at (10.5,0) {};
\draw[dashed] (C2) -- (10.5,-.7);
\draw (6.8,-.7) -- (6.8,-.9) to (9.2,-.9) to (9.2,-.7);
\draw (9.8,-.7) -- (9.8,-.9) to (10.7,-.9) to (10.7,-.7);
\draw[->-=.5] (C2) to (10.2,.35);
\draw[->-=.5] (C1) to (10.2,.35);
\node[scale=.4] at (9.9,.35) {$(2,2)$};
\draw[->-=.7] (B) to (9.6,.7);
\draw (10.2,.35) to (9.6,.7);
\node[scale=.4] at (9.35,.9) {$(4,0)$};
\draw[->-=.5] (9.1,.8) to (A5);
\draw (9.6,.7) to (9.1,.8) to (A5);
\draw (9.1,.8) to [out=170,in=280] (8.3,1.45);
\draw[dashed] (8.3,1.45) to (8.1,2.5);
\draw[->-=.5] (8.1,2.5) to (or1);
\node[scale=.5] at (7.7,2.7) {$(3,0)$};
\node[scale=.5] at (8,-1.1) {A-type};
\node[scale=.5] at (9.5,-1.1) {B-type};
\node[scale=.5] at (10.25,-1.1) {C-type};
\draw[dashed] (8.7,.6) to [out=180,in=90] (6.2,-.55) to [out=270,in=180] (8.7,-1.6) to [out=0,in=270] (11.2,-.55) to [out=90,in=0] (8.7,.6) -- cycle;
\node[scale=1] at (12,.6) {$E_6$ singularity};
\end{scope}
\end{tikzpicture}
\caption{\label{E6}An example of a string network realization of a BPS state. In particular in the figure is shown the highest weight vector of the ${\bf 78}$ with EM charge (3,0).}
\end{figure}

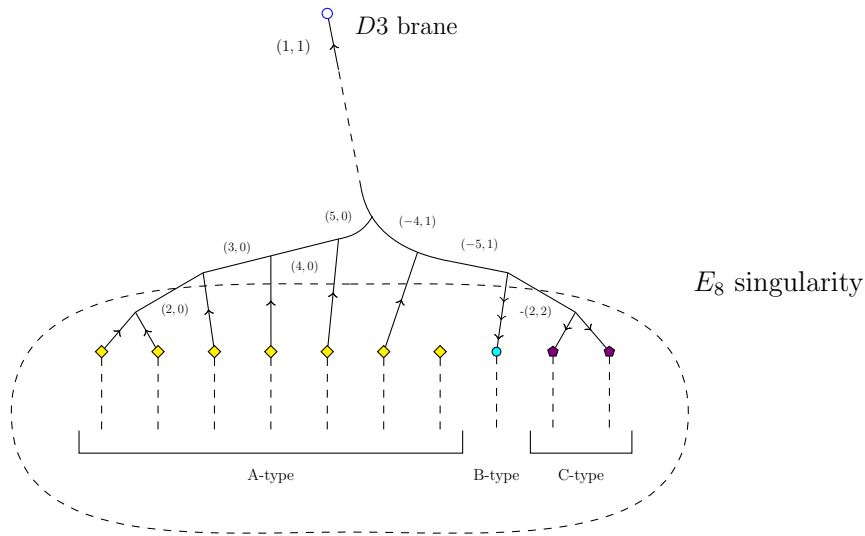
\begin{figure}[tbp]
\centering
\begin{tikzpicture}[decoration={markings,
mark=at position .5 with {\arrow{>}}}]
\begin{scope}[scale=1.5]
\node[Bl,scale=.5] (or1) at (8,3) {};
\node[scale=.8] at (8.7,2.9) {$D3$ brane};
\node[draw,diamond,fill=yellow,scale=.3] (A1) at (6,0) {}; 
\draw[dashed] (A1) -- (6,-.7);
\node[draw,diamond,fill=yellow,scale=.3] (A2) at (6.5,0) {}; 
\draw[dashed] (A2) -- (6.5,-.7);
\node[draw,diamond,fill=yellow,scale=.3] (A3) at (7,0) {}; 
\draw[dashed] (A3) -- (7,-.7);
\node[draw,diamond,fill=yellow,scale=.3] (A4) at (7.5,0) {}; 
\draw[dashed] (A4) -- (7.5,-.7);
\node[draw,diamond,fill=yellow,scale=.3] (A5) at (8,0) {}; 
\draw[dashed] (A5) -- (8,-.7);
\node[draw,diamond,fill=yellow,scale=.3] (A6) at (8.5,0) {}; 
\draw[dashed] (A6) -- (8.5,-.7);
\node[draw,diamond,fill=yellow,scale=.3] (A7) at (9,0) {};
\draw[dashed] (A7) -- (9,-.7);
\node[draw,circle,fill=aqua,scale=.3] (B) at (9.5,0) {};
\draw[dashed] (B) -- (9.5,-.7);
\node[draw,regular polygon,regular polygon sides=5,fill=purple,scale=.3] (C1) at (10,0) {}; 
\draw[dashed] (C1) -- (10,-.7);
\node[draw,regular polygon,regular polygon sides=5,fill=purple,scale=.3] (C2) at (10.5,0) {};
\draw[dashed] (C2) -- (10.5,-.7);
\draw (5.8,-.7) -- (5.8,-.9) to (9.2,-.9) to (9.2,-.7);
\draw (9.8,-.7) -- (9.8,-.9) to (10.7,-.9) to (10.7,-.7);
\draw[->-=.5] (10.2,.35) to (C1);
\draw[->-=.5] (10.2,.35) to (C2);
\node[scale=.4] at (9.85,.35) {-$(2,2)$};
\draw[->-=.8] (9.6,.7) to (9.525,.175);
\draw[->-=.5] (9.525,.175) to (B);
\draw (10.2,.35) to (9.6,.7);
\node[scale=.4] at (9.35,.95) {$(-5,1)$};
\draw[->-=.5] (A6) to (8.8,.88);
\draw (9.6,.7) to (9.1,.8);
\draw (9.1,.8) to [out=170,in=280] (8.3,1.45);
\node[scale=.4] at (8.8,1.15) {$(-4,1)$};
\draw[dashed] (8.3,1.45) to (8.1,2.5);
\draw[->-=.5] (8.1,2.5) to (or1);
\node[scale=.5] at (7.7,2.7) {$(1,1)$};
\draw[->-=.5] (A1) to (6.3,.35);
\draw[->-=.5] (A2) to (6.3,.35);
\draw (6.3,.35) to (6.9,.7);
\node[scale=.4] at (6.65,.37) {$(2,0)$};
\draw[->-=.5] (A3) to (6.9,.7);
\node[scale=.4] at (7.2,.92) {$(3,0)$};
\draw (6.9,.7) to (7.5,.85);
\draw[->-=.5] (A4) to (7.5,.85);
\node[scale=.4] at (7.8,.75) {$(4,0)$};
\draw (7.5,.85) to (8.1,1);
\draw[->-=.5] (A5) to (8.1,1);
\draw (8.1,1) to [out=10, in=240] (8.4,1.2); 
\node[scale=.4] at (8.1,1.2) {$(5,0)$};
\node[scale=.5] at (7.5,-1.1) {A-type};
\node[scale=.5] at (9.5,-1.1) {B-type};
\node[scale=.5] at (10.25,-1.1) {C-type};
\draw[dashed] (8.2,.6) to [out=180,in=90] (5.2,-.55) to [out=270,in=180] (8.2,-1.6) to [out=0,in=270] (11.2,-.55) to [out=90,in=0] (8.2,.6) -- cycle;
\node[scale=.9] at (12,.6) {$E_8$ singularity};
\end{scope}
\end{tikzpicture}
\caption{\label {E8}A slightly more involved example of a string network realization of a BPS state, this time in the $E_8$ theory. In particular in the figure is shown the highest weight vector of the ${\bf 248}$ with EM charge (1,1).}
\end{figure}

The $E_{6,7,8}$ singularity probed by the D3-brane can be resolved by mass
deforming the theory, replacing it by a collection of, respectively, 8, 9 or 10 mutually non-local $(p,q)$ D7-branes. The resulting presentation is not unique; here we use the canonical presentation constructed in \cite{Gaberdiel:1997ud,DeWolfe:1998zf}. This presentation involves multiple branes with charge $(1,0)$, called A-branes, a single brane with charge $(1,-1)$, called B-brane, and two branes of charge $(1,1)$, called C-branes. In this picture all the branes have branch cuts emanating downwards vertically and, from left to right, A-branes appear first, then the B-brane and finally the C-branes; see Figures \ref{E6} and \ref{E8}. What distinguishes the different singularities is $n_{\text{A}}$, the number of A-branes; $n_{\text{A}}=5$ for $E_6$, $n_{\text{A}}=6$ for $E_7$ and $n_{\text{A}}=7$ for $E_8$. For more details see \cite{Gaberdiel:1997ud,DeWolfe:1998zf}.

Recall that, while strings of any charges can end on a D3-brane, only $(\tilde{p},\tilde{q})$-strings can end on a $(\tilde{p},\tilde{q})$-7-brane. Therefore once a presentation of the singularity is given, a string junction can be labeled by a set of integers, \emph{the invariant charges of the junction}, \textbf{Q}$^{\mu} = (Q_A^i,Q_B,Q_C^k)$, $i=1,...,n_{\text{A}}$, $k=1,2$ denoting respectively the signed number of prongs ending on the $n_{\text{A}}$ A-branes, the B-brane and the two C-branes. The asymptotic charge $n(p,q)$ of the junction is then simply given by $np=\sum_{i=1}^{n_{\text{A}}}Q_A^i+Q_B+Q_C^1+Q_C^2$, $nq=Q_C^1+Q_C^2-Q_B$. Of course, the asymptotic charge of the junction does not fully determine the $Q$s; to determine them fully
we also have to specify the flavor charge. 


The invariant charges ${\bf Q}^\mu$ of a string network
are linearly related to the homology class $J$
of the corresponding membrane. Thus henceforth we will use $J$ to indicate interchangeably the homology class of the membrane and the corresponding string network. Furthermore, the intersection number $(J,J')$
can be written as a bilinear in the corresponding invariant charges:
\begin{eqnarray}\nonumber
(J,J')& = -&\sum_{i=1}^{n_{\text{A}}}Q_A^i Q'^i_A-Q_B Q'_B-\sum_{j=1}^2Q_C^j Q'^j_C-\frac{1}{2}\sum_{i=1}^{n_{\text{A}}}Q_A^i(Q'_B-\sum_{j=1}^2Q'^j_C)\\\label{interjunct}
&&-\frac{1}{2} (Q_B-\sum_{j=1}^2Q_C^j)\sum_{i=1}^{n_{\text{A}}}Q'^i_A +Q_B\sum_{j=1}^2 Q'^j_C+\sum_{j=1}^2Q_C^j Q'_B,
\end{eqnarray}
where $n_A=5,6,7$ for $E_{6}$, $E_7$ and $E_8$ respectively. This expression is readily obtained from the property that the self-intersection of a single open string stretching from a single D7 to the D3 is $-1$, and that $(J,J')$ is invariant under continuous deformations of the junctions \cite{DeWolfe:1998zf}.

In \cite{DeWolfe:1998zf} it was observed that there is a natural correspondence between neutral string junction charges which saturate the bounds \eqref{selfint} and the roots of the flavor Lie algebra associated to the given orientifold singularity. 
See \cite{stringweb} for a generalization and purely field-theoretic discussion of this correspondence. 
In a similar fashion we can associate neutral string junction charges $J_{\omega^j}$ to fundamental weights of the Lie algebra, $\vec{\omega}^j$. The $J_{\omega^j}$ satisfy
\begin{equation}
(J_{\alpha_i},J_{\omega^j})=-\delta_i^j.
\label{fundwei}\end{equation}
Two special junction charges can be defined, $J_{\omega_p}$ and $J_{\omega_q}$, which have asymptotic charge $(1,0)$ and $(0,1)$ respectively, and are orthogonal to the fundamental weights, in the sense that:
\begin{equation}
(J_{\omega^j},J_{\omega_p})=(J_{\omega^j},J_{\omega_q})=0.
\label{orto}\end{equation}
\eqref{orto} and the asymptotic charges of $J_{\omega_p}$ and $J_{\omega_q}$ fix these junction charges uniquely. Since the $(J_{\omega_i},J_{\omega_p},J_{\omega_q})$ span the charge lattice it is possible to write any junction charge as
\begin{equation}
J=\sum_{i=1}^r \mu_i J_{\omega^i}+np J_{\omega_p}+nq J_{\omega_q},
\label{charges}\end{equation}
where $r$ is the rank of the flavor Lie algebra.
Since the $J_{\omega_i}$ are neutral it follows that the asymptotic charges of \eqref{charges} are $n(p,q$). Furthermore $(\mu_1, \dots, \mu_r)$ are precisely the Dynkin labels of the flavor representation of the string junction \cite{DeWolfe:1998zf}. From \eqref{fundwei} and \eqref{orto}, the Dynkin labels of a given charge $J$ can be readily extracted:
\begin{equation}
\mu_{i}=-(J,J_{\alpha_{i}}).
\label{dynkin}\end{equation}

Finally, \eqref{dynkin} translates into a relation between the Dynkin label $\mu_i$s and the invariant charges \textbf{Q}$^\mu(J)$. These relations, together with the expression of the EM charges as functions of the $Q$'s, can be inverted. Each junction charge is thus uniquely determined by its Dynkin labels and its asymptotic charges. Since this map is important in our computation, we find it useful to reproduce explicitly, and it is reported in the appropriate section below.

The quadratic form \eqref{interjunct} simplifies considerably if written in terms of the flavor and asymptotic charges of the junctions rather than their invariant charges. In fact, calling $\vec{{\boldsymbol \lambda}}(J)=\sum_i \mu_i \vec{\omega}^i$ the weight vector associated to the junction we have \cite{DeWolfe:1998zf}:
\begin{equation}\label{IntForm}
-(J,J) = \vec{{\boldsymbol \lambda}}(J)\cdot \vec{{\boldsymbol \lambda}}(J)-n^2f(p,q),
\end{equation}
where $f(p,q)$ is a positive definite quadratic form in the asymptotic charges, which varies depending on the flavor symmetry group.

\section{{BPS spectrum in rank \texorpdfstring{$1$}{1} MN theories}}\label{bps_spectrum_of_mn_theories}

We can now compute explicitly the allowed flavor representations for BPS states in rank
$1$ MN theories\footnote{To be precise, for each representation we only consider the junction charge corresponding to the highest weight. The existence of the remaining junctions filling out the entire representation follows by repeatedly adding the neutral string junctions corresponding to negative roots \cite{DeWolfe:1998zf}. In the same paper the authors also counted the number of junctions with self-intersection $-1$ and asymptotic charge $(1,0)$ in the $E_6$, $E_7$ and $E_8$ theories finding 27, 56 and 248 respectively. Thus each fills out the fundamental representation of the flavor group which is obviously consistent with the first row of table \ref{E6n10}, \ref{E7n10} and \ref{E8n10}.}. Below we only present the results for a few $(p,q)$, see table \ref{E6npq}, \ref{E7npq} and \ref{E8npq}, and for $(1,0)$, see table \ref{E6n10}, \ref{E7n10} and \ref{E8n10}. We present the computation for some small values of $n$. It is important to remark that this choice is not due to computational inability; we can quickly and easily compute the allowed flavor representations for very large $(p,q)$ and $n$. Rather, we choose small 
$n$ in order to compare with the existing results in the literature. In labeling the representations we followed the convention in \cite{Yamatsu:2015npn}.

Our results for the $E_6$ theory match perfectly with \cite{Hollands:2016kgm}, and
for the $E_7$ theory they similarly match with \cite{E7bps}. While this is the first time that a calculation of the charges occurring in the BPS spectrum of the $E_8$ theory is carried out, there is an interesting overlap with a recent previous result: our representations in table \ref{E8n10} corresponding to genus 0 curves perfectly match the E-string spectrum of massless particles arising from rational curves of genus 0 recently derived in \cite{Tian:2018icz}. 

Our calculation, despite its remarkable efficiency, is blind to some of the information which can be computed by other methods. In particular, we have no way of computing the index by which a given flavor representation and asymptotic charge appears. This has been done for some small charges in \cite{Hollands:2016kgm,E7bps} for the $E_6$ theory, and in \cite{E7bps} for the $E_7$ theory. Comparing with our results we find ``experimental'' hints of a geometric interpretation of some of the indices, which we
hope can be refined in the future; we will discuss this correspondence in more detail in the next section.

\subsection{\texorpdfstring{$E_6$}{E6} theory}\label{reproducing_e6_theory}

\begin{figure}[tbp]
\centering
\begin{tikzpicture}
\node[draw,circle,fill=white,scale=.75] (r1) at (1,.5) {};
\node at (1,0.15) {$\alpha_1$};
\node[draw,circle,fill=white,scale=.75] (r2) at (2,.5) {};
\node at (2,0.15) {$\alpha_2$};
\node[draw,circle,fill=white,scale=.75] (r3) at (3,.5) {};
\node at (3,0.15) {$\alpha_3$};
\node[draw,circle,fill=white,scale=.75] (r4) at (4,.5) {};
\node at (4,0.15) {$\alpha_4$};
\node[draw,circle,fill=white,scale=.75] (r5) at (5,.5) {};
\node at (5,0.15) {$\alpha_5$};
\node[draw,circle,fill=white,scale=.75] (r6) at (3,1.65) {};
\node at (3.45,1.6) {$\alpha_6$};
\draw[ultra thick] (r1) -- (r2);
\draw[ultra thick] (r2) -- (r3);
\draw[ultra thick] (r3) -- (r4);
\draw[ultra thick] (r4) -- (r5);
\draw[ultra thick] (r3) -- (r6);
\end{tikzpicture}
\caption{Our convention for the labeling of the simple roots of $E_6$.\label{roots_E6}}
\end{figure}
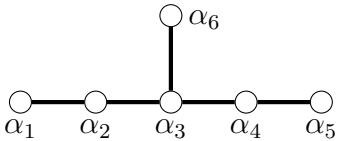

Our convention for the labeling of the $E_6$ simple roots is shown in Figure \ref{roots_E6}, which in turn fixes our convention for the Dynkin labels of the representations. We can then explicitly write the invariant charges of a string junction in terms of its flavor representation Dynkin labels $(\mu_1,...,\mu_6)$ and asymptotic charges $n(p,q)$ \cite{DeWolfe:1998zf}:

\begin{equation}
{\large \boldsymbol{E_6}}:\qquad\left\{\begin{array}{l}
Q^A_1=\frac{1}{3}(4\mu_1+5\mu_2+6\mu_3+4\mu_4+2\mu_5+3\mu_6-np-3nq)\\
Q^A_2=\frac{1}{3}(\mu_1+5\mu_2+6\mu_3+4\mu_4+2\mu_5+3\mu_6-np-3nq)\\
Q^A_3=\frac{1}{3}(\mu_1+2\mu_2+6\mu_3+4\mu_4+2\mu_5+3\mu_6-np-3nq)\\
Q^A_4=\frac{1}{3}(\mu_1+2\mu_2+3\mu_3+4\mu_4+2\mu_5+3\mu_6-np-3nq)\\
Q^A_5=\frac{1}{3}(\mu_1+2\mu_2+3\mu_3+4\mu_4+2\mu_5-np-3nq)\\
Q^B=\frac{1}{3}(-4\mu_1-8\mu_2-12\mu_3-10\mu_4-5\mu_5-6\mu_6+4np-9nq)\\
Q^C_1=\frac{1}{3}(-2\mu_1-4\mu_2-6\mu_3-5\mu_4-\mu_5-3\mu_6+2np-3nq)\\
Q^C_2=\frac{1}{3}(-2\mu_1-4\mu_2-6\mu_3-5\mu_4-4\mu_5-3\mu_6+2np-3nq)
\end{array}\right.
\label{E6map}\end{equation}

\noindent Recall that the $Q^A_i$s, $Q^B$ and $Q^C_{1,2}$ denote respectively the number of prongs ending on the five A-branes, the B-brane and the two C-branes. Plugging \eqref{E6map} into \eqref{interjunct} we find the following expression for the quadratic form in terms of the Dynkin labels and asymptotic charges:
\beq
(J,J)_{E_6}=-\vec{{\boldsymbol \lambda}}(J)\cdot \vec{{\boldsymbol \lambda}}(J)+n^2f_{E_6}(p,q)=-\vec{{\boldsymbol \lambda}}(J)\cdot \vec{{\boldsymbol \lambda}}(J)+n^2\left(\frac13p^2-pq+q^2\right)
\eeq

Now we can set up the computation. First notice that the relations between the ($\mu_i,p,q)$ and the invariant charges of the junctions in \eqref{E6map} involve rational coefficients. Invariant charges can only be realized by an actual junction if they are integers. There do exist choices of Dynkin label and asymptotic charges for which \eqref{E6map} gives invariant charges which are only rational, not integral; we call these \emph{improper charges} (following the terminology of
\cite{DeWolfe:1998zf} where the corresponding would-be junctions were called improper junctions), and we throw them away. We only check \eqref{selfint} and \eqref{mutualint} on \emph{proper charges}.

\begin{table}[tbp]
\centering\small
\begin{tabular}{|c|l|}
\hline
\multicolumn{2}{|c|}{\multirow{2}{*}{\text{\bf Allowed $E_6$ representations for BPS states with $n(1,0)$ asymptotic charge }}}\\
\multicolumn{2}{|c|}{}\\
\hline\hline
$n$ & ($E_6$ representation, $g$) \\ 
\hline
\hline
1&(\textbf{27},0)\\
\hline
2&($\bf{\bar{27}}$,0)\\
\hline
3&(\textbf{78},0)+(\textbf{1},1)\\
\hline
4&(\textbf{351},0)+(\textbf{27},1)\\
\hline
5&($\bf{\bar{1728}}$,0)+($\bf{\bar{351}}$,1)+($\bf{\bar{27}}$,2)\\
\hline
6&($\bf{\bar{5824}}$,0)+(\textbf{2430},0)+(\textbf{2925},1)+(\textbf{650},2)+(\textbf{78},3)+(\textbf{1},4)\\
\hline
7&(\textbf{19305},0)+(\textbf{17550},1)+(\textbf{7371},2)+(\textbf{1728},3)+(\textbf{351$'$},3)+(\textbf{351},4)+(\textbf{27},5)\\
\hline
\multirow{2}{*}{8}&($\bf{\bar{54054}}$,0)+($\bf{\bar{46332}}$,1)+($\bf{\bar{34398}}$,1)+($\bf{\bar{51975}}$,2)+($\bf{\bar{17550}}$,3)+($\bf{\bar{1728}}$,5)+($\bf{\bar{351}'}$,5)\\
&+($\bf{\bar{351}}$,6)+($\bf{\bar{27}}$,7)\\
\hline
\multirow{2}{*}{9}&($\bf{\bar{146432}}$,0)+(\textbf{43758},1)+($\bf{\bar{252252}}$,2)+(\textbf{105600},3)+($\bf{\bar{78975}}$,3)+(\textbf{70070},4)+($\bf{\bar{3003}}$,4)\\
&+(\textbf{34749},5)+(\textbf{5824},6)+($\bf{\bar{5824}}$,6)+(\textbf{2430},6)+(\textbf{2925},7)+(\textbf{650},8)+(\textbf{78},9)+(\textbf{1},10)\\
\hline
\multirow{3}{*}{10}&(\textbf{359424$'$},0)+(\textbf{459459},2)+(\textbf{412776},2)+(\textbf{494208},3)+(\textbf{393822},3)+(\textbf{386100},4)\\
&+(\textbf{61425},4)+(\textbf{314496},5)+(\textbf{112320},6)+(\textbf{46332},6)+(\textbf{34398},6)+(\textbf{51975},7)+(\textbf{19305},7)\\
&+(\textbf{17550},8)+(\textbf{7722},8)+(\textbf{7371},9)+(\textbf{1728},10)+(\textbf{351$'$},10)+(\textbf{351},11)+(\textbf{27},12)\\
\hline
\hline
\end{tabular}
\caption{\label{E6n10}The allowed representations for BPS states with asymptotic charges $n(1,0)$, with $1 \le n \le 10$ are listed. We also keep track of the genus of the corresponding holomorphic curves. We follow \cite{Yamatsu:2015npn} for the $\boldsymbol{R}$, $\boldsymbol{\bar{R}}$ and $\boldsymbol{R}'$ conventions.}
\end{table}

This is the right place to make a quick aside and provide an explanation for a ``superselection'' rule that was noticed in \cite{Hollands:2016kgm}. The compact simply connected form of $E_6$ has a $\mathbb{Z}_3$ center, which acts on a representation with Dynkin label $(\mu_1,...,\mu_6)$ as $\omega^{\mu_1-\mu_2+\mu_4-\mu_5}$, with $\omega=e^{2\pi i/3}$. The authors of \cite{Hollands:2016kgm} pointed out that on all representations allowed for BPS states with asymptotic charge $n(p,q)$, the center acts\footnote{The EM duality frame chosen for the presentation of the $E_6$ singularity in terms of non-mutually local D7-branes in \cite{DeWolfe:1998zf} slightly differs from the choice of duality frame made in \cite{Hollands:2016kgm}. That explains why in \cite{Hollands:2016kgm} the action of the center is written as $\omega^{n(p+q)}$ instead of $\omega^{np}$.} as $\omega^{np}$. In our framework there is a straightforward explanation for this superselection rule: it arises from the requirement that the junction charge is proper. From \eqref{E6map}, the $Q$'s are all of the form $Q_i=\frac13 g_i(\mu_i,n,p,q)$. It follows that the junction is proper iff:
\beq
g_i(\mu_i,n,p,q)\in3\mathbb{Z}, \qquad {\rm i.e.} \qquad \omega^{g_i(\mu_i,n,p,q)}=1.
\eeq 
We can redefine $g_i$ up to multiple of 3 and bring all the $g_i$ to a common form $g(\mu_i,n,p,q)=\mu_1-\mu_2+\mu_4-\mu_5-np \in 3\mathbb{Z}$, which implies 
\beq\label{supE6}
\omega^{\mu_1-\mu_2+\mu_4-\mu_5}\omega^{-np}=1,
\eeq 
from which the superselection rule mentioned above follows.

Let's now go back to our computation. We first fix the asymptotic charge of the string junction, say $n(p,q)$. This fixes the RHS of the inequality in \eqref{selfint}. We then start constructing string junctions with that given asymptotic charge with larger and larger representations. This is done by explicitly plugging different Dynkin labels in \eqref{E6map} along with the chosen $n(p,q)$ and computing the corresponding invariant charges. We then discard the improper junctions. The allowed flavor representations for asymptotic charge $n(p,q)$ are those which give rise to proper junctions that pass both \eqref{selfint} and \eqref{mutualint}.

\begin{table}[tbp]
\centering\small
\begin{tabular}{|c|l|}
\hline
\multicolumn{2}{|c|}{\multirow{2}{*}{\text{\bf $E_6$ representations for BPS states with $n(3,2)$ asymptotic charge }}}\\
\multicolumn{2}{|c|}{}\\
\hline\hline
$n$ & ($E_6$ representation, $g$) \\ 
\hline
\hline
1&(\textbf{78},0)+(\textbf{1},1)\\
\hline
2&(\textbf{650},0)+(\textbf{78},1)+(\textbf{1},2)\\
\hline
3&(\textbf{5824},0)+($\bf{\bar{5824}}$,0)+(\textbf{2925},1)+(\textbf{650},2)+(\textbf{78},3)+(\textbf{1},4)\\
\hline
4&(\textbf{78975},0)+($\bf{\bar{78975}}$,0)+(\textbf{70070},1)+(\textbf{3003},1)+($\bf{\bar{3003}}$,1)+(\textbf{34749},2)+(\textbf{2430},3)\\
&(\textbf{5824},3)+($\bf{\bar{5824}}$,3)+(\textbf{2925},4)+(\textbf{650},5)+(\textbf{78},6)+(\textbf{1},7)\\
\hline
\hline
\end{tabular}
\caption{\label{E6npq}The allowed representations for BPS states with asymptotic charges $n(3,2)$, with $1 \le n \le 4$, are listed. We also keep track of the genus of the corresponding holomorphic curves. We follow \cite{Yamatsu:2015npn} for the $\boldsymbol{R}$ and $\boldsymbol{\bar{R}}$ convention.}
\end{table}

A few remarks are in order. First notice that from \eqref{selfint} we can straightforwardly compute the genus of the putative holomorphic curve associated to a given junction charge (more below). If \eqref{selfint} is saturated, then obviously $g=0$. In the following we will label a BPS state not only by its asymptotic and flavor charge, but also by its genus. We will describe below a correlation between the genus and the \emph{reduced index} of the given flavor representation \cite{Hollands:2016kgm}.

Second, \eqref{mutualint} has to be checked only against charges which do have holomorphic representatives. In practice we start from $n=1$ and then work our way up to larger $n$. As $n$ increases we check \eqref{mutualint} against all the BPS states whose existence has been already established.

The first case we encounter is that of
a single string stretching from one of the D7-branes to the D3; these states have asymptotic charges equal to the charges of a single A, B or C brane. When we go
to larger $n$, at least in the case of $(p,q)=(1,0)$, when \eqref{mutualint} fails, it fails only against this single-string configuration. This is analogous to what happens for the $SU(2)$ $\mathcal{N}=2$ theory \cite{Mikhailov:1997jv}.

Our results are reported in Tables \ref{E6n10} and \ref{E6npq}. The consequences of the superselection \eqref{supE6} are obvious in both tables. For the states in the $(1,0)$ sector \eqref{supE6} reads $\omega^{\mu_1-\mu_2+\mu_4-\mu_5}=\omega^n$ and has a consequence there is a periodicity 3 in the type of representations that appear as $n$ increases. Conversely for the charges in the $(3,1)$ sector, \eqref{supE6} is independent of $n$ and in fact all representations that appears at level $n$, also appear at $n+1$. Moreover we notice that, for the $n(1,0)$ sector, each representation that appears at $(n,g)$ recurs at $(n+3,g+n)$, see table \ref{E7n10}. Similarly in the $(3,1)$ sector, only representations with a trivial action of the center are allowed and the recursion gets modified to $(n,g)\to(n+1,g+n)$.

\begin{table}[tbp]
\centering\small
\begin{tabular}{|c|l|}
\hline
\multicolumn{2}{|c|}{\multirow{2}{*}{\text{\bf Allowed $E_7$ representations for BPS states with $n(1,0)$ asymptotic charge }}}\\
\multicolumn{2}{|c|}{}\\
\hline\hline
$n$ & ($E_7$ representation, $g$) \\ 
\hline
\hline
1&(\textbf{56},0)\\
\hline
2&(\textbf{133},0)+(\textbf{1},1)\\
\hline
3&(\textbf{912},0)+(\textbf{56},1)\\
\hline
4&(\textbf{8645},0)+(\textbf{1539},1)+(\textbf{133},2)+(\textbf{1},3)\\
\hline
5&(\textbf{86184},0)+(\textbf{27664},1)+(\textbf{6480},2)+(\textbf{912},3)+(\textbf{56},4)\\
\hline
\multirow{2}{*}{6}&(\textbf{573440},0)+(\textbf{253935},0)+(\textbf{365750},1)+(\textbf{152152},2)+(\textbf{40755},3)+(\textbf{7371},3)+(\textbf{8645},4)\\
&+(\textbf{1463},4)+(\textbf{1539},5)+(\textbf{133},6)+(\textbf{1},7)\\
\hline
\multirow{2}{*}{7}&(\textbf{3635840},0)+(\textbf{3792096},1)+(\textbf{2282280},2)+(\textbf{861840},3)+(\textbf{320112},3)+(\textbf{362880},4)\\
&+(\textbf{86184},5)+(\textbf{51072},5)+(\textbf{27664},6)+(\textbf{6480},7)+(\textbf{912},8)+(\textbf{56},9)\\
\hline
\hline
\end{tabular}
\caption{\label{E7n10}The allowed representations for BPS states with asymptotic charges $n(1,0)$, with $1 \le n \le 7$ are listed. We also keep track of the genus of the corresponding holomorphic curves.}
\end{table}

\begin{figure}[b!]
\centering
\begin{tikzpicture}
\node[draw,circle,fill=white,scale=.75] (r1) at (1,.5) {};
\node at (1,0.15) {$\alpha_1$};
\node[draw,circle,fill=white,scale=.75] (r2) at (2,.5) {};
\node at (2,0.15) {$\alpha_2$};
\node[draw,circle,fill=white,scale=.75] (r3) at (3,.5) {};
\node at (3,0.15) {$\alpha_3$};
\node[draw,circle,fill=white,scale=.75] (r4) at (4,.5) {};
\node at (4,0.15) {$\alpha_4$};
\node[draw,circle,fill=white,scale=.75] (r5) at (5,.5) {};
\node at (5,0.15) {$\alpha_5$};
\node[draw,circle,fill=white,scale=.75] (r6) at (6,.5) {};
\node at (6,0.15) {$\alpha_6$};
\node[draw,circle,fill=white,scale=.75] (r7) at (3,1.65) {};
\node at (3.45,1.6) {$\alpha_7$};
\draw[ultra thick] (r1) -- (r2);
\draw[ultra thick] (r2) -- (r3);
\draw[ultra thick] (r3) -- (r4);
\draw[ultra thick] (r4) -- (r5);
\draw[ultra thick] (r5) -- (r6);
\draw[ultra thick] (r3) -- (r7);
\end{tikzpicture}
\caption{Our convention for the labeling of the simple roots of $E_7$\label{roots_E7}}
\end{figure}
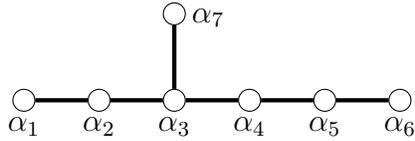

\subsection{\texorpdfstring{$E_7$}{E7} theory}\label{reproducing_known_results_and_new_prediction_for_the_e7_theory}

Our convention for the labeling of the $E_7$ simple roots is shown in fig. \ref{roots_E7}. Again this fixes our convention for the Dynkin labels of the representations, allowing us to write the invariant charges of a string junction in terms of its flavor representation Dynkin labels $(\mu_1,...,\mu_7)$ and asymptotic charges $n(p,q)$ \cite{DeWolfe:1998zf}:

\begin{equation}
{\large \boldsymbol{E_7}}:\qquad\left\{\begin{array}{l}
Q^A_1=\frac{1}{2}(-2\mu_1-4\mu_2-4\mu_3-3\mu_4-2\mu_5-\mu_6-\mu_7-np+3nq)\\
Q^A_2=\frac{1}{2}(-2\mu_1-4\mu_2-4\mu_3-3\mu_4-2\mu_5-\mu_6-3\mu_7-np+3nq)\\
Q^A_3=\frac{1}{2}(-2\mu_1-4\mu_2-6\mu_3-3\mu_4-2\mu_5-\mu_6-3\mu_7-np+3nq)\\
Q^A_4=\frac{1}{2}(-2\mu_1-4\mu_2-6\mu_3-5\mu_4-2\mu_5-\mu_6-3\mu_7-np+3nq)\\
Q^A_5=\frac{1}{2}(-2\mu_1-4\mu_2-6\mu_3-5\mu_4-4\mu_5-\mu_6-3\mu_7-np+3nq)\\
Q^A_6=\frac{1}{2}(-2\mu_1-4\mu_2-6\mu_3-5\mu_4-4\mu_5-3\mu_6-3\mu_7-np+3nq)\\
Q^B=3\mu_1+6\mu_2+8\mu_3+6\mu_4+4\mu_5+2\mu_6+4\mu_7+2p-5nq\\
Q^C_1=2\mu_1+3\mu_2+4\mu_3+3\mu_4+2\mu_5+\mu_6+2\mu_7+np-2nq\\
Q^C_2=\mu_1+3\mu_2+4\mu_3+3\mu_4+2\mu_5+\mu_6+2\mu_7+np-2nq
\end{array}\right.
\label{E7map}\end{equation}
\noindent Plugging \eqref{E7map} into \eqref{interjunct} we find the following expression for the quadratic form in terms of the Dynkin labels and asymptotic charges:
\beq
(J,J)_{E_7}=-\vec{{\boldsymbol \lambda}}(J)\cdot \vec{{\boldsymbol \lambda}}(J)+n^2f_{E_7}(p,q)=-\vec{{\boldsymbol \lambda}}(J)\cdot \vec{{\boldsymbol \lambda}}(J')+n^2\left(\frac12p^2-2pq+\frac52q^2\right)
\eeq

\begin{table}[h!]
\centering\small
\begin{tabular}{|c|l|}
\hline
\multicolumn{2}{|c|}{\multirow{2}{*}{\text{\bf Allowed $E_7$ representations for BPS states with $n(1,1)$ asymptotic charge }}}\\
\multicolumn{2}{|c|}{}\\
\hline\hline
$n$ & ($E_7$ representation, $g$) \\ 
\hline
\hline
1&(\textbf{133},0)+(\textbf{1},1)\\
\hline
2&(\textbf{1539},0)+(\textbf{133},1)+(\textbf{1},2)\\
\hline
3&(\textbf{40755},0)+(\textbf{8645},1)+(\textbf{1463},1)+(\textbf{1539},2)+(\textbf{133},3)+(\textbf{1},4)\\
\hline
\multirow{2}{*}{4}&(\textbf{980343},0)+(\textbf{253935},0)+(\textbf{365750},1)+(\textbf{150822},1)+(\textbf{152152},2)+(\textbf{40755},3)+\\
&(\textbf{7371},3)+(\textbf{8645},4)+(\textbf{1463},4)+(\textbf{1539},5)+(\textbf{133},6)+(\textbf{1},7)\\
\hline
\multirow{3}{*}{5}&(\textbf{23969792},0)+(\textbf{11316305},1)+(\textbf{7482618},1)+(\textbf{7142499},2)+(\textbf{915705},2)+(\textbf{3424256},3)+\\
&(\textbf{617253},3)+(\textbf{573440},4)+(\textbf{980343},4)+(\textbf{253935},4)+(\textbf{365750},5)+(\textbf{150822},5)+\\
&(\textbf{152152},6)+(\textbf{40755},7)+(\textbf{7371},7)+(\textbf{8645},8)+(\textbf{1463},8)+(\textbf{1539},9)+(\textbf{133},10)+(\textbf{1},11)\\
\hline
\hline
\end{tabular}
\caption{\label{E7npq}The allowed representations for BPS states with asymptotic charges $n(1,1)$, with $1 \le n \le 5$ are listed. We also keep track of the genus of the corresponding holomorphic curves. As discussed in the text, only real representations appear.}
\end{table}

The analysis of the allowed BPS spectrum of the $E_7$ MN theory is in large part analogous to the $E_6$. One difference is the ``superselection'' rule on representations which arises from requiring a proper junction for each asymptotic $(p,q)$ charge. The simply connected form of $E_7$ has a $\mathbb{Z}_2$ center which acts on representations with Dynkin label $(\mu_1,...,\mu_7)$ as $(-1)^{\mu_4+\mu_6+\mu_7}$. Analysis similar to the one in the $E_6$ case, shows that \eqref{E7map} gives a proper junction iff:
\beq\label{SuperSelE7}
(-1)^{\mu_4+\mu_6+\mu_7}=(-1)^{n(p+q)}.
\eeq
BPS states with asymptotic charge $n(p,q)$ can only appear in representation satisfy \eqref{SuperSelE7}. The center of $E_7$ acts as $-1$ on pseudoreal representations and as $+1$ on real representations. A consequence of \eqref{SuperSelE7} is thus that in the sector $n(1,0)$, pseudoreal representations appear for $n$ odd and real representations for $n$ even. Moreover we notice that each representation that appears at $(n,g)$ recurs at $(n+2,g+n)$, see table \ref{E7n10}. Similarly, in the $(1,1)$ sector, only real representation are allowed and the recursion gets modified to $(n,g)\to(n+1,g+n)$.

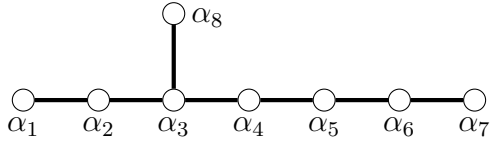
\begin{figure}[tbp]
\centering
\begin{tikzpicture}
\node[draw,circle,fill=white,scale=.75] (r1) at (1,.5) {};
\node at (1,0.15) {$\alpha_1$};
\node[draw,circle,fill=white,scale=.75] (r2) at (2,.5) {};
\node at (2,0.15) {$\alpha_2$};
\node[draw,circle,fill=white,scale=.75] (r3) at (3,.5) {};
\node at (3,0.15) {$\alpha_3$};
\node[draw,circle,fill=white,scale=.75] (r4) at (4,.5) {};
\node at (4,0.15) {$\alpha_4$};
\node[draw,circle,fill=white,scale=.75] (r5) at (5,.5) {};
\node at (5,0.15) {$\alpha_5$};
\node[draw,circle,fill=white,scale=.75] (r6) at (6,.5) {};
\node at (6,0.15) {$\alpha_6$};
\node[draw,circle,fill=white,scale=.75] (r7) at (7,.5) {};
\node at (7,0.15) {$\alpha_7$};
\node[draw,circle,fill=white,scale=.75] (r8) at (3,1.65) {};
\node at (3.45,1.6) {$\alpha_8$};
\draw[ultra thick] (r1) -- (r2);
\draw[ultra thick] (r2) -- (r3);
\draw[ultra thick] (r3) -- (r4);
\draw[ultra thick] (r4) -- (r5);
\draw[ultra thick] (r5) -- (r6);
\draw[ultra thick] (r6) -- (r7);
\draw[ultra thick] (r3) -- (r8);
\end{tikzpicture}
\caption{Our convention for the labeling of the simple roots of $E_8$\label{roots_E8}}
\end{figure}

\begin{table}[bp]
\centering\small
\begin{tabular}{|c|l|}
\hline
\multicolumn{2}{|c|}{\multirow{2}{*}{\text{\bf Allowed $E_8$ representations for BPS states with $n(1,0)$ asymptotic charge }}}\\
\multicolumn{2}{|c|}{}\\
\hline\hline
$n$ & ($E_8$ representation, $g$) \\ 
\hline
\hline
1&(\textbf{248},0)+(\textbf{1},1)\\
\hline
2&(\textbf{3875},0)+(\textbf{248},1)+(\textbf{1},2)\\
\hline
3&(\textbf{147250},0)+(\textbf{30380},1)+(\textbf{3875},2)+(\textbf{248},3)+(\textbf{1},4)\\
\hline
\multirow{2}{*}{4}&(\textbf{6696000},0)+(\textbf{2450240},1)+(\textbf{779247},2)+(\textbf{147250},3)+(\textbf{27000},3)+(\textbf{30380},4)+\\
&(\textbf{3875},5)+(\textbf{248},4)+(\textbf{1},7)\\
\hline
\multirow{3}{*}{5}&(\textbf{301694976},0)+(\textbf{146325270},1)+(\textbf{76271625},2)+(\textbf{26411008},3)+(\textbf{4881384},3)+\\
&(\textbf{6696000},4)+(\textbf{4096000},4)+(\textbf{2450240},5)+(\textbf{779247},6)+(\textbf{147250},7)+(\textbf{27000},7)+\\
&(\textbf{30380},8)+(\textbf{3875},9)+(\textbf{248},10)+(\textbf{1},11)\\
\hline
\multirow{5}{*}{6}&($\bf{8634368000'}$,0)+(\textbf{4076399250},0)+(\textbf{6899079264},1)+(\textbf{4825673125},2)+\\
&(\textbf{2275896000},3)+(\textbf{820260000},3)+(\textbf{1094951000},4)+(\textbf{203205000},4)+\\
&(\textbf{344452500},5)+(\textbf{301694976},5)+(\textbf{146325270},6)+(\textbf{70680000},6)+(\textbf{76271625},7)+\\
&(\textbf{1763125},7)+(\textbf{26411008},8)+(\textbf{4881384},8)+(\textbf{6696000},9)+(\textbf{4096000},9)+\\
&(\textbf{2450240},10)+(\textbf{779247},11)+(\textbf{147250},12)+(\textbf{27000},12)+(\textbf{30380},13)+(\textbf{3875},14)+\\
&(\textbf{248},15)+(\textbf{1},16)\\
\hline
\hline
\end{tabular}
\caption{\label{E8n10}The allowed representation for BPS states with asymptotic charges $n(1,0)$, with $1 \le n \le 6$ are listed. We also keep track of the genus of the corresponding holomorphic curves.}
\end{table}

\subsection{$E_8$ theory}\label{bps_spectrum_of_the_e8_theory}

Our convention for the labeling of the $E_8$ simple roots is shown in fig.~\ref{roots_E8}. As before, this fixes our convention for the Dynkin label of the representations which allows us to write the invariant charges of a string junction in terms of its flavor representation Dynkin labels $(\mu_1,...,\mu_8)$ and asymptotic charges $n(p,q)$ \cite{DeWolfe:1998zf}:

\begin{equation}
{\large \boldsymbol{E_8}}:\qquad\left\{\begin{array}{l}
Q^A_1=-2\mu_1-4\mu_2-5\mu_3-4\mu_4-3\mu_5-2\mu_6-\mu_7-2\mu_8-np+3nq\\
Q^A_2=-2\mu_1-4\mu_2-5\mu_3-4\mu_4-3\mu_5-2\mu_6-\mu_7-3\mu_8-np+3nq\\
Q^A_3=-2\mu_1-4\mu_2-6\mu_3-4\mu_4-3\mu_5-2\mu_6-\mu_7-3\mu_8-np+3nq\\
Q^A_4=-2\mu_1-4\mu_2-6\mu_3-5\mu_4-3\mu_5-2\mu_6-\mu_7-3\mu_8-np+3nq\\
Q^A_5=-2\mu_1-4\mu_2-6\mu_3-5\mu_4-4\mu_5-2\mu_6-\mu_7-3\mu_8-np+3nq\\
Q^A_6=-2\mu_1-4\mu_2-6\mu_3-5\mu_4-4\mu_5-3\mu_6-\mu_7-3\mu_8-np+3nq\\
Q^A_7=-2\mu_1-4\mu_2-6\mu_3-5\mu_4-4\mu_5-3\mu_6-2\mu_7-3\mu_8-np+3nq\\
Q^B=7\mu_1+14\mu_2+20\mu_3+16\mu_4+12\mu_5+8\mu_6+4\mu_7+10\mu_8+4np-11nq\\
Q^C_1=4\mu_1+7\mu_2+10\mu_3+8\mu_4+6\mu_5+4\mu_6+2\mu_7+5\mu_8+2np-5nq\\
Q^C_2=3\mu_1+7\mu_2+10\mu_3+8\mu_4+6\mu_5+4\mu_6+2\mu_7+5\mu_8+2np-5nq
\end{array}\right.
\label{E8map}\end{equation}

\noindent Plugging \eqref{E8map} into \eqref{interjunct} we find the following expression for the quadratic form in terms of the the Dynkin labels and asymptotic charges:
\beq
(J,J)_{E_8}=-\vec{{\boldsymbol \lambda}}(J)\cdot \vec{{\boldsymbol \lambda}}(J)+n^2f_{E_8}(p,q)=-\vec{{\boldsymbol \lambda}}(J)\cdot \vec{{\boldsymbol \lambda}}(J')+n^2\left(p^2-5pq+7q^2\right)
\eeq

Notice that in the $E_8$ case the {\it improper charges} cannot arise. In fact the map \eqref{E8map} only involves integer coefficients with the consequence that, for any $(p,q)$, each representation that occurs at $n$ recurs at $n+1$. The result for the $n(1,0)$ charge sector, up to $n=6$, is tabulated in the table \ref{E8n10} above. For $n=1$, only the trivial and adjoint representations occur (at genus $g=1,0$, respectively). In this case, each representation occuring at $(n,g)$, recurs at $(n+1,g+n)$. In particular, the \emph{highest} genus associated to a given value of $n$ (the one at which the trivial representation occurs) is $g(n)= 1+\frac12 n(n+1)$. In addition, \emph{new} representations occur, for each $n$, at $g\leq 2n-5$. The occurrence of new representations and recurrence of old ones as a function of $n$ and $g$ is particularly evident in table \ref{E8different} below. In table \ref{E8npq} we instead list the representations appearing in the $n(1,1)$ sector up to $n=3$.

\begin{table}[h!]
\centering\small
\begin{tabular}{|c|c|c|c|c|c|c|}
\hline
$g\backslash n$&1&2&3&4&5&6\\
\hline
\multirow{3}{*}{0}&$\multirow{3}{*}{\color{red}248}$&$\multirow{3}{*}{\color{red}3875}$&$\multirow{3}{*}{\color{red}147250}$&$\multirow{3}{*}{\color{red}6696000}$&$\multirow{3}{*}{\color{red}301694976}$&$\color{red}8634368000'$\\[-3pt]
&&&&&&+\\[-3pt]
&&&&&&${\color{red}4076399250}$\\
\hline
1&$\color{red}1$&$248$&$\color{red}30380$&$\color{red}2450240$&$\color{red}146325270$&$\color{red}6899079264$\\
\hline
2&&$1$&$3875$&$\color{red}779247$&$\color{red}76271625$&$\color{red}4825673125$\\
\hline
\multirow{3}{*}{3}&&&\multirow{3}{*}{$248$}&$147250$&$\color{red}26411008$&$\color{red}2275896000$\\[-3pt]
&&&&+&+&+\\[-3pt]
&&&&$\color{red}27000$&$\color{red}4881384$&$\color{red}820260000$\\
\hline
\multirow{3}{*}{4}&&&$\multirow{3}{*}{1}$&$\multirow{3}{*}{30380}$&$6696000$&$\multirow{3}{*}{\color{red}1094951000}$\\[-3pt]
&&&&&+&\\[-3pt]
&&&&&$\color{red}4096000$&\\
\hline
\multirow{3}{*}{5}&&&&\multirow{3}{*}{$3875$}&\multirow{3}{*}{$2450240$}&$\color{red}344452500$\\[-3pt]
&&&&&&+\\[-3pt]
&&&&&&301694976\\
\hline
\multirow{3}{*}{6}&&&&\multirow{3}{*}{$248$}&\multirow{3}{*}{779247}&$146325270$\\[-3pt]
&&&&&&+\\[-3pt]
&&&&&&$\color{red}70680000$\\
\hline
\multirow{3}{*}{7}&&&&\multirow{3}{*}{$1$}&147250&76271625\\[-3pt]
&&&&&+&+\\[-3pt]
&&&&&27000&$\color{red}1763125$\\
\hline
\multirow{3}{*}{8}&&&&&\multirow{3}{*}{$30380$}&26411008\\[-3pt]
&&&&&&+\\[-3pt]
&&&&&&4881384\\
\hline
$\multirow{3}{*}{9}$&&&&&$\multirow{3}{*}{3875}$&6696000\\[-3pt]
&&&&&&+\\[-3pt]
&&&&&&4096000\\
\hline
10&&&&&$248$&$2450240$\\
\hline
11&&&&&$1$&$779247$\\
\hline
\multirow{3}{*}{12}&&&&&&147250\\[-3pt]
&&&&&&+\\[-3pt]
&&&&&&27000\\
\hline
13&&&&&&$30380$\\
\hline
14&&&&&&$3875$\\
\hline
15&&&&&&$248$\\
\hline
16&&&&&&$1$\\
\hline
\end{tabular}
\caption{\label{E8different} We report a different way of looking of the allowed representation for BPS states of the $E_8$ theory with asymptotic charges $n(1,0)$. The representation is indicated in red when it occurs for the first time.}
\end{table}

\subsection{Symmetries of the BPS spectrum}\label{Symmetries_BPS}

It was pointed out in \cite{DeWolfe:1998bi} that the BPS
spectra of the MN theories are invariant under a certain cyclic group action on the
charges. It is worth recalling the argument here.  

\begin{table}[tbp]
\centering\small
\begin{tabular}{|c|l|}
\hline
\multicolumn{2}{|c|}{\multirow{2}{*}{\text{\bf Allowed $E_8$ representations for BPS states with $n(1,1)$ asymptotic charge}}}\\
\multicolumn{2}{|c|}{}\\
\hline\hline
$n$ & ($E_8$ representation, $g$) \\ 
\hline
\hline
1&(\textbf{3875},0)+(\textbf{248},1)+(\textbf{1},2)\\
\hline
\multirow{2}{*}{2}&(\textbf{2450240},0)+(\textbf{779247},1)+(\textbf{147250},2)+(\textbf{27000},2)+(\textbf{30380},3)+(\textbf{3875},4)+\\
&(\textbf{248},5)+(\textbf{1},6)\\
\hline
\multirow{4}{*}{3}&(\textbf{281545875},0)+(\textbf{2275896000},0)+(\textbf{820260000},0)+(\textbf{203205000},1)+\\
&(\textbf{1094951000},1)+(\textbf{344452500},2)+(\textbf{301694976},2)+(\textbf{146325270},3)+(\textbf{70680000},3)+\\
&(\textbf{76271625},4)+(\textbf{1763125},4)+(\textbf{26411008},5)+(\textbf{4881384},5)+(\textbf{6696000},6)+\\
&(\textbf{4096000},6)+(\textbf{2450240},7)+(\textbf{779247},8)+(\textbf{147250},9)+(\textbf{27000},9)+(\textbf{30380},10)+\\
&(\textbf{3875},11)+(\textbf{248},12)+(\textbf{1},13)\\
\hline
\hline
\end{tabular}
\caption{\label{E8npq}The allowed representation for BPS states with asymptotic charges $n(1,1)$, with $1\geq n\geq3$ are listed. We also keep track of the genus of corresponding holomorphic curve.}
\end{table}

The set of low-energy EM charges ($p,q$) (here $(p,q)$ are not necessarily mutually prime) carries an action of the low-energy EM duality group $SL(2,\Z)\cong \Sp(2,\Z)$. The bilinear form naturally defined on the string junctions  can be written as a sum of two parts \eqref{IntForm}, the $SL(2,\Z)$ group only acts on the quadratic form $f(p,q)$. It is then natural to ask whether there exists a non-trivial subgroup of the low-energy EM duality group which leaves $f(p,q)$ invariant. From the way in which we have studied the BPS states in this paper, it is obvious that if such a subgroup exists, the allowed flavor representations of the BPS states associated to any $(p,q)$ charges related by these transformations will be identical. The authors of \cite{DeWolfe:1998bi} carried out this analysis for the $E_6$, $E_7$ and $E_8$ theories, finding that $f(p,q)$ is left invariant respectively by a $\Z_6$, $\Z_4$ and $\Z_6$ subgroup of $SL(2,\Z)$. We can also write down the explicit expression for them (we are using the same notation as \cite{DeWolfe:1998bi}):
\begin{itemize}
\item[$\boldsymbol{E_6}$:] 
\begin{equation}
M^0_{\pm}(6)=
\left(
\begin{array}{cc}
\pm1&0\\
0&\pm1
\end{array}
\right),\quad
M^1_{\pm}(6)=
\left(
\begin{array}{cc}
\pm1&\mp3\\
\pm1&\mp2
\end{array}
\right),\quad
M^2_{\pm}(6)=
\left(
\begin{array}{cc}
\mp2&\pm3\\
\mp1&\pm1
\end{array}
\right).
\end{equation}

\item[$\boldsymbol{E_7}$:]
\begin{equation}
M^0_{\pm}(7)=
\left(
\begin{array}{cc}
\pm1&0\\
0&\pm1
\end{array}
\right),\quad
M^1_{\pm}(7)=
\left(
\begin{array}{cc}
\pm2&\mp5\\
\pm1&\mp2
\end{array}
\right).
\end{equation}

\item[$\boldsymbol{E_8}$:]
\begin{equation}
M^0_{\pm}(8)=
\left(
\begin{array}{cc}
\pm1&0\\
0&\pm1
\end{array}
\right),\quad
M^1_{\pm}(8)=
\left(
\begin{array}{cc}
\pm2&\mp7\\
\pm1&\mp3
\end{array}
\right),\quad
M^2_{\pm}(8)=
\left(
\begin{array}{cc}
\pm3&\mp7\\
\pm1&\mp2
\end{array}
\right).
\end{equation}
\end{itemize}
As it was already pointed out in \cite{DeWolfe:1998bi}, $M(6)_-^0$, $M(6)_-^1$ and $M(6)_-^2$ have to be accompanied by a non-trivial action of the outer automorphism of $E_6$, and thus the allowed representations for BPS states with charges $M(6)_-^{0,1,2}(p,q)$ will be conjugated compared to the allowed representations for states with charge $(p,q)$. 

These expected symmetries of the spectrum can be checked explicitly in our computations and we find that the in the $E_6$ case the allowed representations for $n(1,0)$, $n(1,1)$ and $n(-2,-1)$, at fixed $n$, are identical. For $E_7$ the relevant charges are $n(1,0)$ and $n(2,1)$, and for $E_8$ they are $n(1,0)$, $n(2,1)$ and $n(3,1)$.

\section{Geometry and properties of the BPS states}\label{geometry_and_properties_of_the_bps_states}

As discussed throughout the manuscript, our computation is very efficient in computing allowed flavor representations, but it is insensitive to their indices. We also already mentioned that each BPS state is not only characterized by its EM charge and flavor representation, but also by two integers, $(n,g)$: $g$ is the genus and $n$ the number of boundaries of the holomorphic curve associated to it. Here we collect some ``experimental'' results which suggest a geometrical interpretation of some of the BPS indices. By ``experimental'' we mean that we did not have an \textit{a priori} reason to expect that the quantities described below should be related; we were guided instead by our computation and the comparison with existing results.

\subsection{Divisibility and the number of boundaries}\label{purity}

Let us recall a bit of background on BPS 1-particle states of $\cN=2$ theories. 
For a fixed electromagnetic charge and momentum, the BPS 1-particle 
Hilbert space is
a representation of  $SU(2)_{\rm spin}\times SU(2)_R$,
of the form
$ \cH = \left( (1/2,0)\oplus(0,1/2) \right) \otimes \cH_{\rm red}$,
where the ``reduced Hilbert space'' $\cH_{\rm red}$ is also a representation of
$SU(2)_{\rm spin}\times SU(2)_R$.
The {\it no-exotic} conjecture, proposed in \cite{Gaiotto:2010be}, states that 
the action of $SU(2)_R$ on $\cH_{\rm red}$ is trivial.
This conjecture has been proven in some particular cases \cite{Chuang:2013wt,DelZotto:2014bga}, and a general proof will appear in \cite{CD}.
There still remains the question of describing $\cH_{\rm red}$, either 
computing its dimension or more ambitiously to determine its 
$SU(2)_{\rm spin}$ representation content. This is what we usually mean by
the question of ``determining the BPS multiplicities.''

Fortunately there are BPS indices available which are well suited to this
problem. The simplest such index is the second helicity supertrace, 
which is simply the graded dimension (superdimension) 
of $\cH_{\rm red}$. There is also the protected spin character
introduced in \cite{Gaiotto:2010be}, which keeps track of the $SU(2)_{\rm spin}$
content of $\cH_{\rm red}$, which is strictly more information than the 
graded dimension.

As of now there are no computations of the protected spin character available
in Minahan-Nemeschansky theories. However, there are some computations of
the second helicity supertrace, which paint a suggestive picture.
In particular, it was observed in \cite{Hollands:2016kgm} that in the 
rank $1$
Minahan-Nemeschansky $E_6$ and $E_7$ theories, BPS states with charge $n(p,q)$ occur with
second helicity supertrace divisible by $(-1)^{n+1} n$. The same property continues to hold in the $E_7$ theory \cite{E7bps}. This extra divisibility was not
predicted in advance, and seems to be pointing to some additional structure
in these theories.

Since $(-1)^{n+1} n$ is exactly the contribution
to the second helicity supertrace from a spin-$\frac{n}{2}$ multiplet of 
$SU(2)_{\rm spin}$, it seems natural to conjecture that $\cH_{\rm red} = V_n \otimes W$ where $V_n$ is the spin-$\frac{n}{2}$
multiplet, and $W$ is some representation of $SU(2)_{\rm spin}$; then
the second helicity supertrace would just be 
$\sdim \cH_{\rm red} = (-1)^{n+1} n \, \sdim W$,
which is indeed divisible by $(-1)^{n+1} n$, as long as $\sdim W > 0$.
(An even stronger conjecture would be that $SU(2)_{\rm spin}$ acts trivially
on $W$; this was called the ``spin purity'' hypothesis in \cite{Hollands:2016kgm}.)

The F-theory perspective suggests a nice geometric interpretation of this factorization of $\cH_{\rm {red}}$. As already discussed in \S\ref{string_junctions_and_bps_states}, $n$, the greatest common divisor of the asymptotic charges, corresponds to the number of boundaries of the holomorphic curves associated to a given BPS state with charge $n(p,q)$. This is true, regardless of the flavor representation under which the state transforms. 
It is then appealing to conjecture that the quantization of M2-branes with $n$ boundaries
generally gives rise to a Hilbert space including a spin-$\frac{n}{2}$ multiplet
as a universal factor. The fact that the number of boundaries of the holomorphic curve wrapped by an M2-brane is related to the spin of the resulting BPS states was already noticed for the BPS hypermultiplet and vector multiplet, respectively realized as holomorphic curves with one boundary (disk) and two boundaries (cylinder) \cite{Henningson:1997hy,Mikhailov:1997jv}.

If this is the correct interpretation of the divisibility property of 
BPS indices in rank $1$ MN theories, 
then we should expect that this property will be modified in more general
theories. For example, we could consider the rank $r$ MN theories.
In this case the picture of BPS states is similar to rank $1$, except that the 
M2-branes can have boundary components on any of $r$ distinct torus fibers. For a state with
electromagnetic charge $(n_1(p_1,q_1), n_2(p_2,q_2), \dots, n_r(p_r,q_r))$
	the number of boundaries is $\sum_{i=1}^r n_i$. Thus the natural analogue of 
	the divisibility conjecture in this case would be that there is still
	a universal factor $V_n$ in the BPS Hilbert space,
	but now $n$ is not just the GCD of the charges: rather $n = \sum_{i=1}^r n_i$.

\subsection{Reduced indices and genuses}

As discussed above, the evidence in  \cite{Hollands:2016kgm,E7bps} supports
the hypothesis that the second helicity supertraces of BPS states of 
the $E_6$ and $E_7$ theories with charge $n(p,q)$ are always 
integer multiples of $(-1)^{n+1} n$. It is then natural to consider a {\it reduced} index \cite{Hollands:2016kgm} by dividing by $(-1)^{n+1} n$.

While the number of boundaries of the holomorphic curve representing a given BPS state only depends on the EM charges, the genus depends on the flavor charge as well. This is easy to see from \eqref{holcur} and \eqref{IntForm}:
\beq\label{gen}
g_{\vec{{\boldsymbol \lambda}}(J),n,p,q}=\frac{n^2f(p,q)-n+1-\vec{{\boldsymbol \lambda}}(J)\cdot\vec{{\boldsymbol \lambda}}(J)}{2},
\eeq
where $f(p,q)$ is a positive definite quadratic form. \eqref{gen} implies that the genus is a monotonically decreasing function of the norm of the weight vector of the BPS state. For a given EM charge $n(p,q)$, BPS states in the singlet representation, if allowed, will be always represented by holomorphic curves with the maximal genus.

We notice that in all cases in which a BPS state is represented by a holomorphic curve with genus $0$ and information on the corresponding reduced index is available, the reduced index is exactly $1$. This leads us to conjecture that holomorphic curves with genus $0$ and $n$ boundaries in a given relative homology class are not only isolated but unique, and they only give rise to a single BPS multiplet, of spin $\frac{n}{2}$.

We also notice that the reduced index of BPS states increases with the genus. In particular, BPS states associated to holomorphic curves with genus $1$ always appear either with reduced index $2$ or $3$. From genus $2$ onwards, the range of reduced indices which can appear increases rapidly. The central value of the range is a monotonically increasing function of the genus. It would be interesting to better understand where in the geometry the information about the indices is encoded. Our preliminary analysis strongly suggests that the genus of the holomorphic curve is part of the story, but to have a more complete picture more information is needed.






\acknowledgments

It is a pleasure to thank Philip Argyres, Qianyu Hao and David Morrison for useful discussions. JD was supported by  NSF grant PHY-1620610. MM was supported in part by NSF grant PHY-1151392 and in part by NSF grant PHY-1620610.  AN was supported by NSF grant DMS-1711692.
Part of this work was performed at the Aspen Center for Physics, which is supported by National Science Foundation grant PHY-1607611.

\bibliographystyle{JHEP}
\bibliography{BPSEtheories.bib}

\end{document}